\documentclass[doublecol,figures]{epl2}
\usepackage{amssymb,amsmath}
\usepackage{epsf,colordvi,graphicx,color,amsbsy}
\usepackage[T1]{fontenc}

\title{Origin of intermittency in wave turbulence}
\author{E. Falcon\inst{1}\thanks{Corresponding author: \email{eric.falcon@univ-paris-diderot.fr}} \and S. G. Roux\inst{2} \and C. Laroche\inst{1}} 
\shortauthor{E. Falcon, S. G. Roux \and C. Laroche}
\institute{
\inst{1} Laboratoire Mati\`ere et Syst\`emes Complexes (MSC), Universit\'e Paris Diderot, CNRS -- UMR 7057\\ 10 rue A. Domon \& L. Duquet, 75 013 Paris, France, EU\\
\inst{2} Laboratoire de Physique, Ecole Normale Sup\'erieure de Lyon, CNRS -- UMR 5672, 69 007 Lyon, France, EU
}
 
\pacs{47.35.+i}{Hydrodynamic waves} 
\pacs{47.35.Jk}{Wave breaking} 
\pacs{47.27.-i}{Turbulent flows} 
\abstract{Using standard signal processing tools, we experimentally report that intermittency of wave turbulence on the surface of a fluid occurs even when two typical large-scale coherent structures (gravity wave breakings and bursts of capillary waves on steep gravity waves) are not taken into account. We also show that intermittency depends on the power injected into the waves. The dependence of the power-law exponent of the gravity-wave spectrum on the forcing amplitude cannot be also ascribed to these coherent structures. Statistics of these both events are studied.} 

\begin{document}

\maketitle
Understanding the origin of intermittency is a challenging problem in varied domains involving turbulent flows. Intermittency is the occurrence of bursts of intense motion within more quiescent fluid flow \cite{Batchelor49,K62}. This leads to strong deviations from Gaussian statistics that become larger and larger when considering fluctuations at smaller and smaller scales. In three-dimensional hydrodynamic turbulence, the origin of these deviations has been ascribed to the formation of coherent structures (strong vortices) since the 50's \cite{Batchelor49}. However, the physical mechanism of intermittency is still an open question \cite{Arneodo08}. Intermittency has also been observed in granular systems \cite{EPL}, in magnetohydrodynamic turbulence in geophysics \cite{DeMichelis04} or in the solar wind \cite{Olga}, and in systems involving transport by a turbulent flow \cite{Falkovitch01}. A recent observation of intermittency  has been reported in wave turbulence  \cite{Falcon07interm}, a system that strongly differs from high Reynolds number hydrodynamic turbulence. It could thus motivate explanations of intermittency different than the ones considering the dynamics of the Navier-Stokes equation. 

The aim of this Letter is to investigate if some coherent structures are responsible of intermittency in wave turbulence. In the case of wave turbulence on a surface of a fluid, coherent structures such as bursts of capillary waves on steep gravity waves \cite{Cox58} and wave breakings \cite{Duncan99} are well-known phenomena, these latters being recently taken into account in numerical simulations \cite{Zakharov07}. Wave breakings also occur in plasma waves, internal waves, and Rossby waves in geophysics. It has been suggested that intermittency in wave turbulence may be connected to wave structures (such as cusps, whitecaps or wave breakings) thus motivating theoretical \cite{Newell} and numerical \cite{Yokoyama04} works. Here, we show experimentally that intermittency does not come from wave breakings and capillary bursts on gravity waves. Using standard signal processing tools, one finds criteria to detect such structures that allow us to study their statistics and their possible role in the origin of intermittency. We also show that intermittency depends on the power injected into the waves. The frequency-power law of the gravity-wave spectrum is known to depend on the forcing parameters \cite{Falcon07}. We show that this dependence is not related to these coherent structures.

The experimental setup has been described previously \cite{Falcon07}. It consists of a square vessel, 20 $\times$ 20 cm$^2$ filled with mercury up to a height of 2.6 cm. Similar results are found with water. Surface waves are generated by the horizontal motion of a rectangular plunging plastic wave maker driven by an electromagnetic exciter. This vibration exciter is driven with a random forcing within a narrow low-frequency range (typically 0.1 to 5 Hz), and a rms voltage amplitude $\sigma_U$ from 0.1 to 0.8 V leading to wave mean steepnesses (ratio of crest-to-trough amplitude to its duration) from 1 up to 4 cm/s. The rms value $\sigma_V$ of the velocity fluctuations of the wave maker is proportional to $\sigma_U$. The mean power injected $\langle I \rangle$ into the fluid scales as $\langle I \rangle \sim \sigma^2_V \sim \sigma_U^2$ \cite{Falcon08}. $\varepsilon \equiv \langle I \rangle/(\rho A)$ is the mean energy flux where $\rho$ is the fluid density and $A$ the immersed area of the wave maker. The surface wave amplitude, $\eta(t)$, is measured at a given location of the surface by a capacitive wire gauge \cite{Falcon07}. $\eta(t)$ is low-pass filtered at 1 kHz and recorded with a 4 kHz sampling rate during 3000 s, leading to $N=1.2 \times 10^7$ points. This signal is cut in 10 files of 300~s. Statistical properties of each file are then computed, the rms value of the computed quantity giving its error bar. This also allows us to check the signal stationarity. 

\begin{figure}[!t]
\epsfysize=60mm
\epsffile{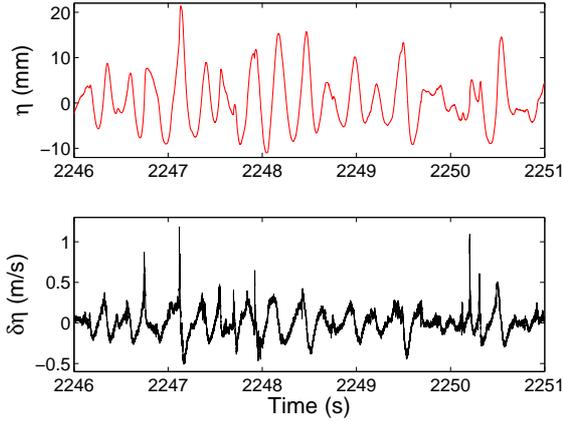}
\caption{(color online) Temporal evolution of the surface wave amplitude (top) and the local slope of the wave (bottom) for a strong forcing ($\sigma_U=0.8$ V) during 5 s.}
\label{fig01}
\end{figure}

For a weak forcing, the wave amplitude $\eta(t)$ is found to fluctuate around a zero mean value in a roughly Gaussian way, as well as the local slope of the surface waves $\delta_\eta(t)$ computed from the differential of $\eta(t)$. For a higher forcing amplitude, a typical temporal recording of the wave amplitude is displayed in Fig.\ \ref{fig01}: $\eta(t)$ much more fluctuates with more probable high crest waves than deep trough waves. This comes from nonlinear effects due to the strong steepness of the waves. The corresponding local slope $\delta_\eta(t)$ is also strongly erratic (see Fig.\ \ref{fig01}), and two typical events can be observed: short peaks of very high amplitudes, and trains of oscillations of much smaller amplitudes  (see Fig.\ \ref{fig01} and below). Both events occur randomly, and are always found close to the maximum of the local slope of the wave. 

\begin{figure}[!t]
\centerline{
\begin{tabular}{c}
\epsfysize=50mm
\epsffile{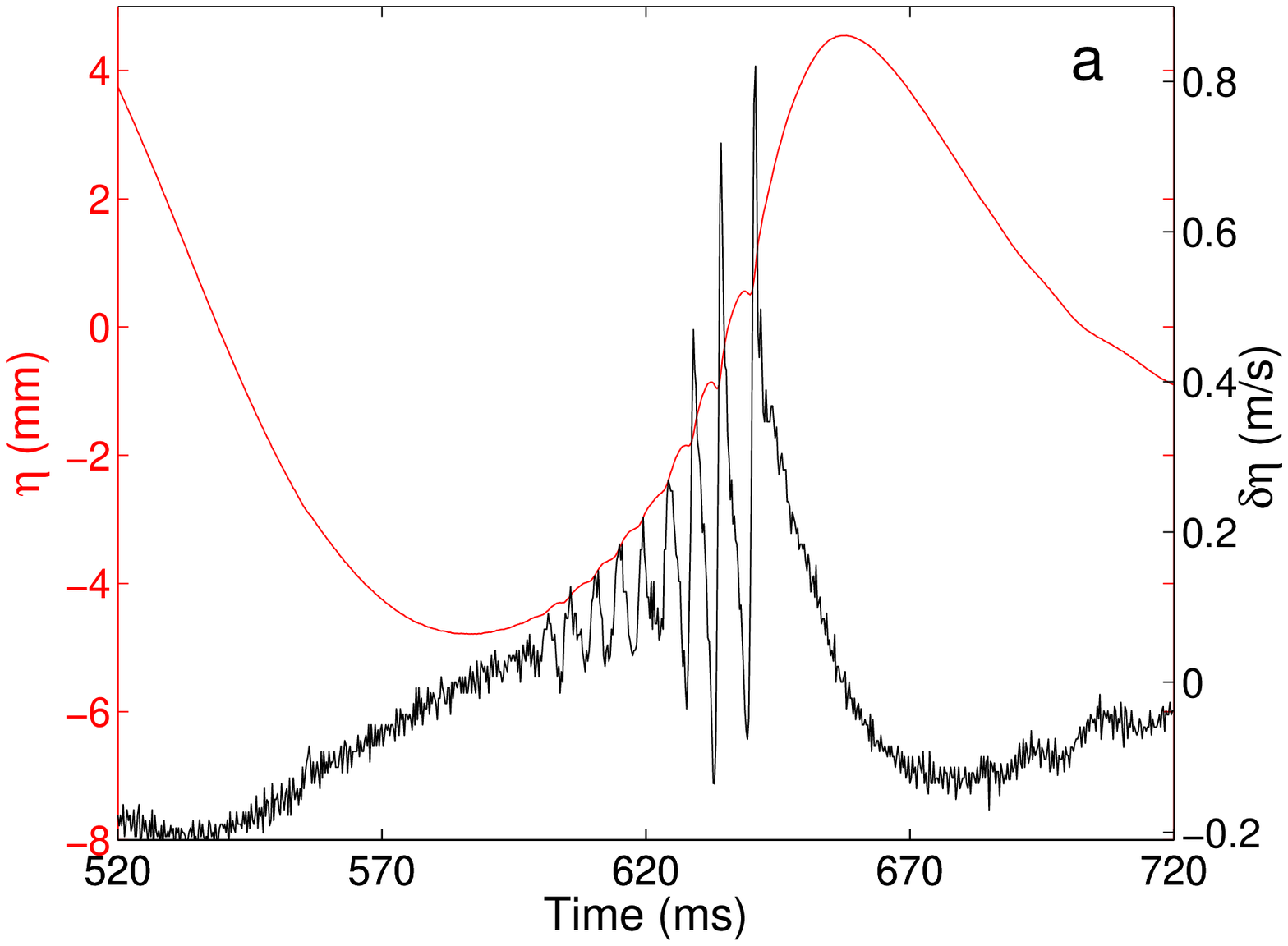}\\
\epsfysize=50mm
\epsffile{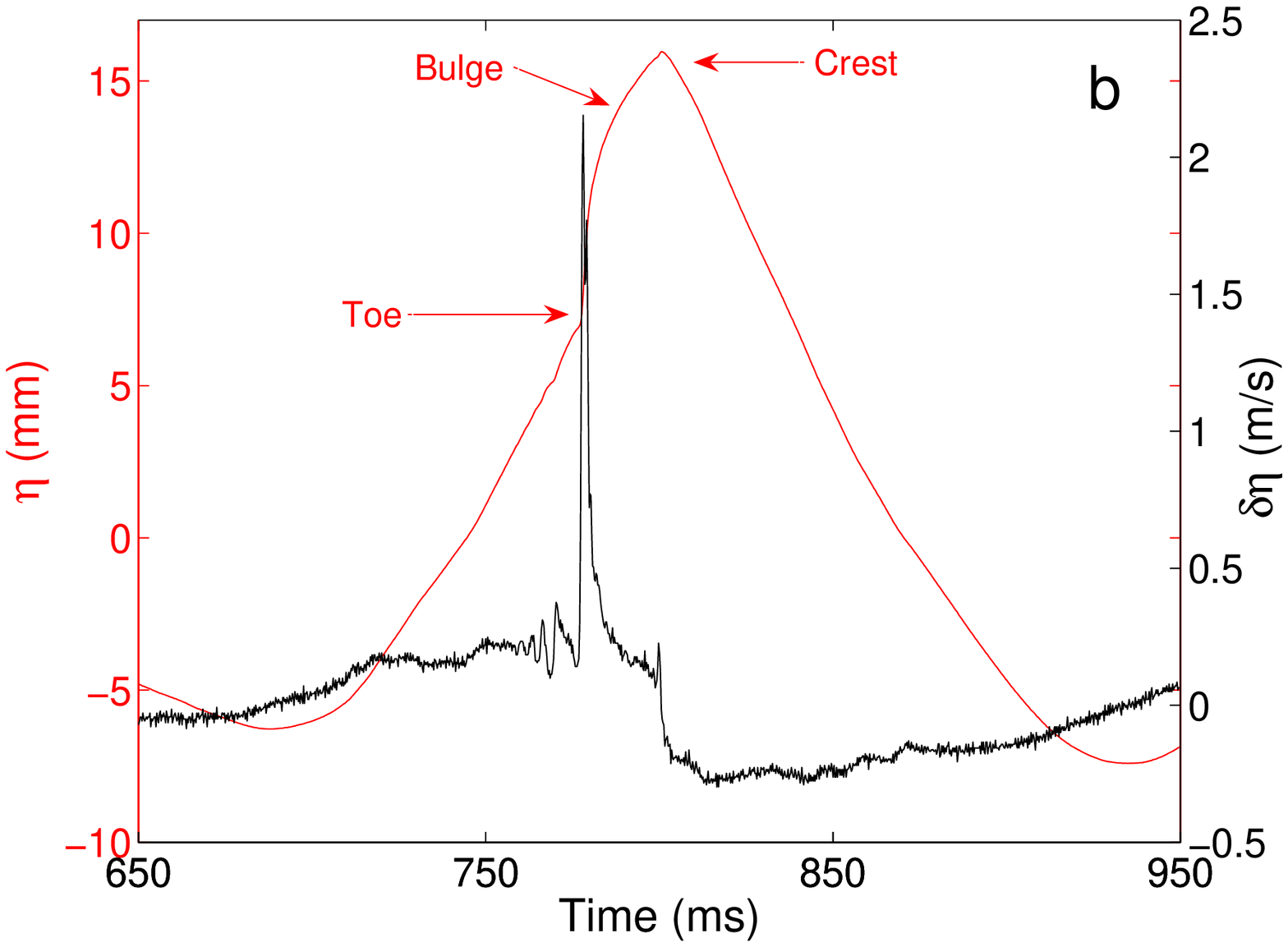}
\end{tabular}
}
\caption{(color online) Typical events occurring near the crest of a steep gravity wave: {\bf a)} Capillary waves generation ($\sigma_U=0.4$ V). {\bf b)} Wave breaking ($\sigma_U=0.8$ V). Both events occur on the forward face of the gravity wave (wave front is left-hand side). Axis of the wave amplitude (resp. local slope) is on the left-hand side (resp. right-hand side). Upper (resp. lower) curve corresponds to the wave amplitude (resp. wave slope).}
\label{fig02}
\end{figure}

Figure\ \ref{fig02} shows such typical events that are detected on the wave amplitude signal once the forcing is high enough. Figure\ \ref{fig02}a shows the first kind of event: a burst of high-frequency capillary waves on a gravity wave. This is a well-known phenomenon occurring when the gravity wave is steep enough \cite{Cox58}. Indeed, when the gravity wave amplitude increases, the local curvature at the crest increases rapidly leading to strong surface-tension effects close to the crest. A train of capillary waves then is emitted propagating down the gravity wave front face as predicted theoretically \cite{LH63}. One can also observe in Fig.\ \ref{fig02}a that the wave is much steeper in the front than in the rear. The capillary wave frequency (obtained from the period between two successive peaks in the slope trace) is found to increase with the index of number of capillary waves from the gravity wave crest as already shown experimentally \cite{Cox58} and predicted theoretically  \cite{LH63}. The typical frequency of the carrier gravity wave is of the order of 5 Hz, whereas the capillary wave ones are in the range $80 - 250$ Hz. 

Figure\ \ref{fig02}b shows the second type of event observed at high enough forcing. Sharp peaks occur on the wave-slope signal, corresponding to the early stage of a wave breaking: a ``bulge'' is formed on the forward face near the crest preceded by small-amplitude capillary waves. It is well-known that as the wave steepens, the amplitude of the bulge increases \cite{Duncan99}. The leading edge of the bulge (also called the ``toe'') marks the formation of a train of small-amplitude capillary waves. These capillary waves generally grow rapidly with time leading to the breaking of the wave \cite{Duncan99}. 

\begin{figure}[!t]
\epsfysize=65mm
\epsffile{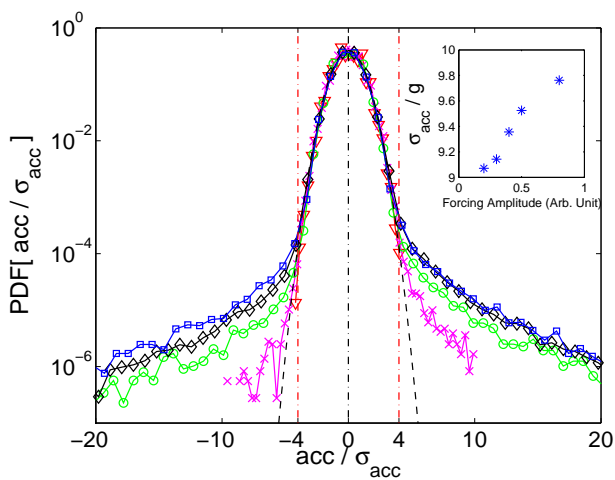}
\caption{(Color online) PDF of the normalized wave acceleration, $acc/\sigma_{acc}$, for different forcings $\sigma_U=$ 0.2 ($\triangledown$), 0.3 ($\times$), 0.4 ($\circ$), 0.5 ($\diamond$) and 0.8 ($\square$) V  (from lower to upper curves). Gaussian with zero mean and unit standard deviation (black dashed line). Wave breaking onset is $\pm4\sigma_{acc}$ [red (light gray) dot-dashed lines]. Inset: $\sigma_{acc}$ as a function of the forcing.}
\label{fig03}
\end{figure}

As shown in Fig.\ \ref{fig02}, a wave breaking has a larger slope amplitude than the one of a burst of ripples generated on the gravity wave. Consequently, one can find a criterion to detect wave-breaking events in order to study their statistics. To wit, the local acceleration of the surface waves is computed from the second differential of $\eta(t)$. The probability density function (PDF) of the acceleration, $acc$, normalized to its rms value, $\sigma_{acc}$, is shown in Fig.\ \ref{fig03} for different forcing amplitudes. The core of the PDF is Gaussian and independent of the forcing up to a critical wave acceleration of $\pm 4\sigma_{acc}$. Above this value, the PDF tails depend on the forcing: the larger the forcing, the more probable the rare events are. This critical acceleration is the onset of wave breakings. Indeed, as directly observed on the acceleration signal, wave breakings occur when $acc \gtrsim 4\sigma_{acc}$. One can thus remove from the acceleration statistics the wave-breaking events, i.e. a set of short signal durations $\delta t\equiv t_f-t_i$ where the absolute value of the acceleration becomes larger at time $t_i$ (resp. lower at $t_f$) than this threshold (typically $\delta t\simeq 100$ ms). When removing these wave-breaking events, the PDF of the filtered acceleration is then found to be almost Gaussian whatever the forcing. Note that this should not be confused with a lack of intermittency (see below). The inset of Fig.\ \ref{fig03} shows that $\sigma_{acc}$ increases with the forcing as expected.

\begin{figure}[!t]
\epsfysize=70mm
\epsffile{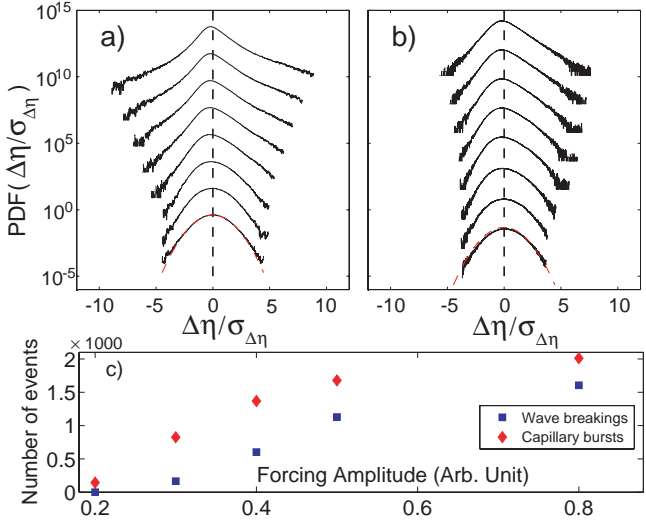} 
\caption{(Color online) PDF of normalized increments $\Delta\eta / \sigma_{{\Delta}\eta}$ for different time lags $\tau=$ 20, 23, 27, 32, 38, 51, 70, 132 ms (from top to bottom) computed from: {\bf a)} whole signal, {\bf b)} signal without wave breakings and capillary burst events. Gaussian fit (dashed line). Curves have been shifted for clarity. $\sigma_U=0.3$~V. {\bf c)} Number of detected events {\it vs.} the forcing.}
\label{fig04}
\end{figure}

Statistics of wave-breakings is then performed using the above detection criterion. Figure \ \ref{fig04}c shows the number of wave breaking detected as a function of the forcing. For the smallest forcing, no wave breaking occurs on the fluid surface. When the forcing is increased, the number of wave breakings increases. Note that the PDF of time lag between two consecutive wave-breaking events is found to exponentially decrease as expected for Poissonian statistics (not shown here). The PDF of a number of events occurring in a fixed period of time is also found to follow a Poisson law with an occurrence rate of events increasing with the forcing.

Let us now focus on the detection of capillary bursts on gravity waves. A time-frequency analysis based on wavelet transforms is a useful method to analyse signals with multiple time-varying frequencies \cite{Flandrin}. It provides temporal and spectral information simultaneously and is thus well adapted to detect capillary burst events by thresholding the energy (i.e., the wavelet coefficients squared) contains in the bandwidth $50$ -- $250$ Hz. One thus obtains a set of signal durations $\delta t=t_f-t_i$ where the wavelets coefficients are above the threshold. $\delta t$ is of the order of 40 to 80 ms from the smallest to the strongest forcing. Figure \ \ref{fig04}c shows the number of capillary bursts detected by this process as a function of the forcing. When the forcing is increased, the number of capillary bursts is found to increase. We also find that the statistics of capillary bursts does not follow a Poissonian distribution.

Wave breaking and capillary burst events are two different coherent structures that can independently be detected within the signal $\eta(t)$ (see above). One can thus probe their respective role in the  intermittency phenomenon in wave turbulence. The intermittent properties of a stochastic stationary signal are generally tested by computing the structure functions using the first-order differences of the signal $\delta\eta(\tau)\equiv \eta(t+\tau)-\eta(t)$. However, when the signal has a steep power spectrum, $E_{\eta}(f) \sim f^{-n}$ with $n>3$, high-order difference statistics is required \cite{Falcon07interm}. For gravity wave turbulence, the theoretical exponent of the power spectrum of wave amplitude is $n=4$ \cite{Zakharov67Grav}, whereas experimentally it is found to depend on the forcing with $n\geq 4$ \cite{Falcon07}, the origin of the discrepancy being an open problem. We found that statistical convergence of the structure functions is reached when using the fourth-order (or higher) difference statistics. This is due to our  locally multi-derivable signal \cite{Muzy93}. The fourth-order differences of the signal
 ${\Delta}\eta_t(\tau)\equiv {\eta}(t+2\tau)-4{\eta}(t+\tau)+6{\eta}(t)-4{\eta}(t-\tau)+{\eta}(t-2\tau)$, are thus computed in the following.
 
The PDFs of ${\Delta}\eta_t(\tau)$ normalized to their rms values $\sigma_{{\Delta}\eta}$ are computed, for different time lags $20\leq \tau \leq 130$ ms, either from the whole signal $\eta(t)$ (see Fig.\ \ref{fig04}a) or from the signal where both wave breaking and capillary burst events are removed (see Fig.\ \ref{fig04}b).  In both cases, a shape deformation of the PDFs is observed with $\tau$. The PDF is roughly Gaussian at large $\tau$, and its shape changes continuously when $\tau$ is decreased. This is a direct signature of intermittency \cite{Pope}. Since this latter is observed in both cases (Fig.\ \ref{fig04}a and \ref{fig04}b), it clearly means that wave breaking and capillary burst events are not the origin of intermittency.
%Finally, we have checked that the PDF of ${\Delta}\eta(\tau)$ still undergoes a shape deformation across the scales $\tau$ when removing the coherent structures. 

\begin{figure}[t]
\centerline{
\epsfysize=64mm
\epsffile{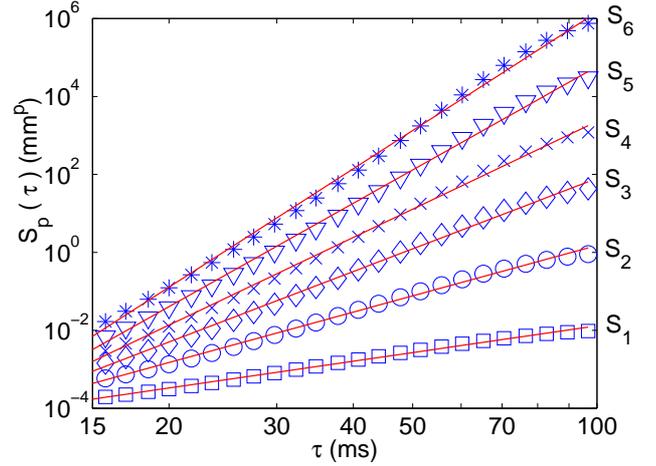}
}
\caption{(Color online) Structure functions ${\mathcal S}_p(\tau)$ of the fourth-order differences of the wave amplitude, $\Delta\eta$, as functions of the time lag $\tau$, for $1\leq p \leq 6$ (as labeled). Solid line: Power law fits, ${\mathcal S}_p \sim \tau^{\xi_p}$, where the slopes $\xi_p$ depend on the order $p$ (see Fig.\ \ref{fig06}). Curves have been shifted for clarity.  $\sigma_U=0.3$ V. }
\label{fig05}
\end{figure}

\begin{figure}[t]
\centerline{
\epsfysize=65mm
\epsffile{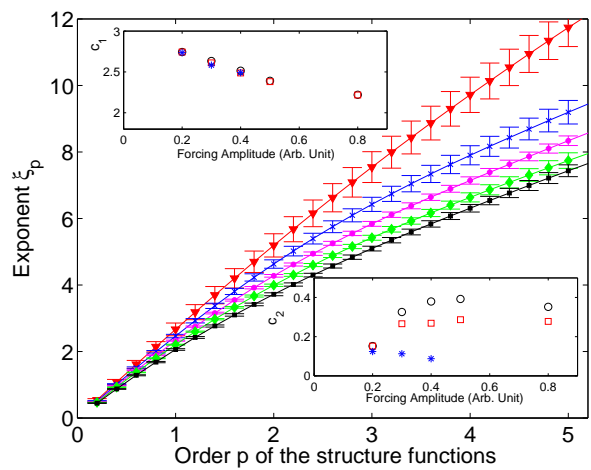}
}
\caption{(Color online) Exponents $\xi_p$ of the structure functions as a function of $p$ for different forcings: $\sigma_U=$ ($\triangledown$) 0.2 , ($\times$) 0.3, ($\circ$) 0.4, ($\diamond$) 0.5  and ($\square$) 0.8 $V$  (from upper to lower curves). Solid lines are best fits $\xi_p=c_1p-\frac{c_2}{2}p^2$. $\xi_p$ are computed from the fourth-order differences with $15 \leq \tau \leq 100$ ms ({\it e.g.}, see Fig.\ \ref{fig05}). Top and bottom insets: evolution of $c_1$ and $c_2$ with the forcing: ($\circ$) whole signal, ($\square$) signal without wave breakings, and ($\ast$) signal without wave breakings and capillary bursts (for the three lowest forcings). }
\label{fig06}
\end{figure}

The structure functions  are defined by
\begin{equation}
{\mathcal S}_p(\tau)\equiv  \langle |\Delta\eta_t(\tau)|^p \rangle=\frac{1}{N}\sum_{t=1}^{N} |\Delta\eta_t(\tau)|^p \; ,
\label{eq:sf}
\end{equation} 
where N is the total number of points in the signal. ${\mathcal S}_p(\tau)$ are shown in Fig.\ \ref{fig05} for a fixed forcing. For $15\leq \tau \leq 100$ ms, all the structure functions of order $p$ (from 1 to 6) are found to be power laws of $\tau$: ${\mathcal S}_p(\tau) \sim \tau^{\xi_p}$ where $\xi_p$ is an increasing function of the order $p$. The exponents $\xi_p$ of the structure functions are then plotted in the main Fig.\ \ref{fig06} as a function of $p$ for different forcings. $\xi_p$ is found to be a nonlinear function of $p$ such that $\xi_p=c_1p-\frac{c_2}{2}p^2$ where the values of $c_1$ and $c_2$ are found to both depend on the forcing (see $\circ$-symbols in top and bottom insets of Fig.\ \ref{fig06}). The nonlinearity of $\xi_p$ ($c_2\neq 0$) is a second signature of intermittency \cite{Pope}. The so-called intermittency coefficient $c_2$ is found to increase from 0.2 up to 0.4 when the forcing is increased whereas $c_1$ is found to decrease from 2.8 to 2.2. Intermittency is observed here over almost one decade in time ($15\leq \tau \leq 100$ ms), corresponding to frequencies $5 \leq 1/(2\tau) \leq 33$ Hz related to gravity wave turbulence regime.  Indeed, as observed on the power spectrum of the wave amplitudes (not shown here), this upper boundary value is the crossover frequency between gravity and capillary wave turbulence regimes for the driving frequency bandwidth used. This crossover is known to depend on the forcing parameters \cite{Falcon07}, and is thus slightly increased with respect to the one between linear gravity and capillary waves $\frac{1}{\sqrt{2}\pi}\left(\frac{g^3\rho}{\gamma}\right)^{1/4}$ ($\simeq 17$ Hz for mercury).

From the definition of the structure functions of Eq.\ (\ref{eq:sf}), one can remove the coefficients $\Delta\eta_t(\tau)$ obtained at times $t$  in a given neighborough of a wave-breaking or a capillary-burst event, i.e. for $t\in [t_i-T,t_f+T]$.  We choose $T=160$ ms to ensure that all kept coefficients are not polluted by the event. Moreover, $T$ has to be larger than the maximum time scale (100 ms) used for the estimation of $\xi_p$. The drawback of this method is that a lot of statistics is removed just for a single event. Using this signal processing protocol, one can thus performed a structure function analysis by removing either all the wave-breaking events detected ($\square$-symbols in the insets of Fig.\ \ref{fig06}), or by removing both the wave breaking and capillary burst events ($\ast$-symbols in the insets) from the statistics in Eq. (\ref{eq:sf}).  When wave breakings are removed, the intermittency coefficient $c_2$ is found to slightly decreases for a fixed forcing, but  $c_2$ is still non zero whatever the forcing (see $\square$-symbols in the bottom inset). This means that intermittency is still observed, and thus cannot be ascribed to the wave breakings. For a fixed forcing, when both wave breaking and capillary burst events are removed, the $c_2$ strongly decreases ($\ast$-symbols), but it is still above 0.1, that is, one order of magnitude larger than the typical values found in usual hydrodynamic turbulence \cite{Pope}. Capillary bursts thus enhance intermittency but are not its primary origin. We have checked that these results do not depend on the signal processing used. When changing T strongly (from 160 to 650 ms), intermittency is still observed ($c_2$ decreases slightly from 0.13 to 0.11 for the lowest forcing). This means that although from 15\% to 40\% of the original signal is removed, intermittency is very robust and could not be ascribed to possible missing events of the signal processing. To sum up, both PDF and structure function analyses lead to coherent results showing that intermittency occurs even when two typical coherent structures (wave breakings and capillary bursts) are not taken into account.

Finally, let us now focus on the frequency exponent of the gravity wave spectrum that is related to the second order structure function by $n=\xi_2+1=2(c_1-c_2)+1$. From the values of $c_1$ and $c_2$, our measurements show a decrease of $n$ from $6.2$ to 4.6 for an increasing forcing. When removing the coherent structures from the signal, the value of $c_1$ is found to be independent of the presence of these both events (see top inset of Fig.\ \ref{fig06}), and the exponent $n$ of the gravity spectrum, estimated from $n=2(c_1-c_2)+1$, is still found to be a decreasing function of the forcing amplitude. Coherent structures are thus not the origin of the forcing-dependent frequency-power law spectrum. Such a forcing-dependent exponent is coherent with previous direct observations of the gravity wave spectrum \cite{Falcon07,Denissenko07}. This departure from the theoretical spectrum of gravity wave $E^{theo}_{\eta}(f) \sim \varepsilon^{1/3}gf^{-4}$ (with $g$ the acceleration of the gravity and $\varepsilon$ the mean energy flux) \cite{Zakharov67Grav} could be related to finite size effect of the container and/or to strong nonlinear effect of high wave steepness in experiments \cite{Falcon07,Denissenko07}. Indeed, weak turbulence theory assumes infinite basin and weak nonlinearity \cite{Zakharov67Grav}. For this weak regime, a simple dimensional analysis thus leads to ${\mathcal S}^{theo}_p(\tau) \sim \varepsilon^{p/6}g^{p/2}\tau^{3p/2}$. Here, one finds experimentally, for small p, ${\mathcal S}_p(\tau) \sim \varepsilon^{\alpha p}\tau^{c_1p}$ where $\alpha=0.4\pm 0.05$ and $c_1$ depends on the forcing (see inset of Fig.\ \ref{fig06}).

In conclusion, we have shown that intermittency in wave turbulence persists when two typical large-scale coherent structures (wave breakings and capillary bursts on steep gravity waves) are removed from the wave amplitude signal. The power-law exponent of the gravity-wave spectrum is known to be non-universal in laboratory experiments \cite{Falcon07,Denissenko07}. Here, we show that this exponent dependence on the forcing parameters cannot be also ascribed to these coherent structures. The origin of the intermittency phenomenon in wave turbulence is still an open issue. It could be ascribed to the large fluctuations of the energy flux \cite{Falcon08} or to other possible wave structures. 

\acknowledgments
We thank B. Audit for discussion. This work has been supported by ANR turbonde BLAN07-3-197846.

%%%%%%%%%%%%%%%%%%%%%%%%%%%%%%%%%%%%%%
%%%%%%%%%%%% REFERENCES %%%%%%%%%%%%%%%%%%
%%%%%%%%%%%%%%%%%%%%%%%%%%%%%%%%%%%%%%


\begin{thebibliography}{99}
\bibitem{Batchelor49}\Name{Batchelor G. K. \and Townsend A. A.} \REVIEW{Proc. R. Soc. Lond. A}{199}{1949}{238}
\bibitem{K62}\Name{Kolmogorov A. N.} \REVIEW{J. Fluid Mech.}{13}{1962}{82}
\bibitem{Arneodo08}\Name{Arn\'eodo A. \etal} \REVIEW{Phys. Rev. Lett.}{100}{2008}{254504}; \Name{Chevillard L. \and Meneveau C.} \REVIEW{C. R. M\'ecanique}{335}{2007}{187}; \REVIEW{Phys. Rev. Lett.}{97}{2006}{174501}
\bibitem{EPL}\Name{Falcon E., Castaing B. \and Laroche C.} \REVIEW{Europhys. Lett.}{65}{2004}{186}
\bibitem{DeMichelis04}\Name{De Michelis P. \and Consolini G.} \REVIEW{Ann. Geophys.}{47}{2004}{1713} 
\bibitem{Olga}\Name{Alexandrova O., Carbone V., Veltri P., Sorriso-Valvo L.} \REVIEW{Planetary and Space Science}{55}{2007}{2224} 
\bibitem{Falkovitch01}\Name{Falkovitch G., Gawedzki K. \and Vergassola M.} \REVIEW{Rev. Mod. Phys.}{73}{2001}{913}
\bibitem{Falcon07interm}\Name{Falcon E., Fauve S. \and Laroche C.}\REVIEW{Phys. Rev. Lett.}{98}{2007}{154501}
\bibitem{Cox58}\Name{Cox C. S.} \REVIEW{J. Mar. Res.}{16}{1958}{199};  \Name{Chang J. H., Wagner R. N. and Yuen H. C.} \REVIEW{J. Fluid Mech.}{86}{1978}{401}
\bibitem{Duncan99}\Name{Duncan J. H., Qiao H., Philomin V. \and  Wenz A.} \REVIEW{J. Fluid Mech.}{379}{1999}{191}, and references therein.
\bibitem{Zakharov07}\Name{Zakharov V. E., Korotkevich A. O., Pushkarev A. \and Resio D.} \REVIEW{Phys. Rev. Lett.} {99}{2007}{164501}; \Name{Korotkevich A. O., Pushkarev A., Resio D. \and  Zakharov V. E.} \REVIEW{Eur. J. Mech. B/Fluids} {27}{2008}{361}.
\bibitem{Newell}\Name{Connaughton C., Nazarenko S. \and Newell A. C.} \REVIEW{Physica D}{184}{2003}{86}; \Name{Biven L. , Nazarenko S. \and Newell A.} \REVIEW{Phys. Lett. A}{280}{2001}{28}
\bibitem{Yokoyama04}\Name{Yokoyama N.} \REVIEW{J. Fluid Mech.}{501}{2004}{69}
\bibitem{Falcon07}\Name{Falcon E., Laroche C. \and Fauve S.} \REVIEW{Phys. Rev. Lett.}{98}{2007}{094503}
\bibitem{Falcon08}\Name{Falcon  E.  \etal} \REVIEW{Phys. Rev. Lett.} {100}{2008}{064503}.
\bibitem{LH63}\Name{Longuet-Higgins M. S.} \REVIEW{J. Fluid Mech.} {16}{1963}{138}
\bibitem{Flandrin}\Name{Flandrin P.} \Book{Time-Frequency/Time-Scale Analysis} \Publ{Academic Press Inc., London} \Year{1999}.
\bibitem{Zakharov67Grav}\Name{Zakharov V. E. \and Filonenko N. N.} \REVIEW{Phys. Dokl.}{11}{1967}{881}
\bibitem{Muzy93}\Name{Muzy J.-F., Bacry E. \and Arn{\'e}odo A.} \REVIEW{Phys. Rev. E}{47}{1993}{875}
\bibitem{Pope}\Name{Pope S. B.} \Book{Turbulent flows} \Publ{Cambridge University, Cambridge} \Year{2006}.
\bibitem{Denissenko07}\Name{Denissenko P., Lukaschuk S. \and Nazarenko S.} \REVIEW{Phys. Rev. Lett.}{99}{2007}{014501}
%\bibitem{Nazarenko09}S. Lukaschuck, S. Nazarenko, S. McLelland and P. Dennissenko, Phys. Rev. Lett {\bf 103}, 044501 (2009).
\end{thebibliography}
\end{document}